\documentclass[fleqn,usenatbib]{mnras}

\usepackage{newtxtext,newtxmath}

\usepackage[T1]{fontenc}

\DeclareRobustCommand{\VAN}[3]{#2}
\let\VANthebibliography\thebibliography
\def\thebibliography{\DeclareRobustCommand{\VAN}[3]{##3}\VANthebibliography}

\usepackage{graphicx}	
\usepackage{amsmath}
\title[Candidate PeVatron  MGRO~J1908+06]{Multiwavelength investigation of the candidate Galactic PeVatron MGRO~J1908+06}

\author[S. Crestan et al.]{
S. Crestan$^{1,2}$\thanks{E-mail: silvia.crestan@inaf.it},
A. Giuliani$^{2}$,
S. Mereghetti$^{2}$,
L. Sidoli$^{2}$,
F. Pintore$^{3}$,
N. La Palombara$^{2}$
\\
$^{1}$ Università degli Studi dell'Insubria, Via Valleggio 11, 22100 Como, Italy\\
$^{2}$ INAF -- IASF Milano, Via A. Corti 12, 20133 Milano, Italy \\
$^{3}$ INAF -- IASF Palermo, Via U. La Malfa 153, 90146 Palermo, Italy \\
}

\date{Accepted XXX. Received YYY; in original form ZZZ}

\pubyear{2020}

\begin{document}
\label{firstpage}
\pagerange{\pageref{firstpage}--\pageref{lastpage}}
\maketitle

\begin{abstract}
The candidate PeVatron MGRO~J1908+06, which shows a hard spectrum beyond 100~TeV, is one of the most peculiar $\gamma$-ray sources in the Galactic plane. Its complex morphology and some possible counterparts spatially related with the VHE emission region, preclude to distinguish between a hadronic or leptonic nature of the $\gamma$-ray emission.
In this paper we illustrate a new multiwavelength analysis of MGRO~J1908+06, with the aim to shed light on its nature and the origin of its ultra high-energy emission.
We performed an analysis of the $^{12}$CO and $^{13}$CO molecular line emission  demonstrating the presence of dense molecular clouds spatially correlated with the source region. We also analyzed   12-years of \textit{Fermi-LAT} data between 10~GeV and 1~TeV finding a counterpart with a hard spectrum ($\Gamma \sim 1.6$). Our reanalysis of \textit{XMM-Newton} data allowed us to put a more stringent constraint on the X-ray flux from this source. 
We demonstrate that a single accelerator cannot explain the whole set of multiwavelength data, regardless of whether it accelerates protons or electrons, 
but a 2-zone model is needed to explain the emission from MGRO~J1908+06. 
The VHE emission seems most likely the superposition of a TeV PWN powered by PSR~J1907+0602, in the southern part, and of the interaction between the supernova remnant G40.5-0.5 and the molecular clouds towards the northern region.
\end{abstract}


%%%%%%%%%%%%%%%%% BODY OF PAPER %%%%%%%%%%%%%%%%%%
\begingroup
\let\clearpage\relax

\endgroup
\section{Introduction}
Cosmic rays (CRs) are   high-energy atomic nuclei (90\% protons) moving through space at nearly the speed of light. 
The local CR proton spectrum is well described by a power-law up to the so-called ‘knee’ around 1~PeV ($= 10^{15}$~eV), indicating the existence of powerful proton accelerators (‘PeVatrons’) residing in our Galaxy. However, despite decades of efforts, no specific Galactic source has been securely identified as a proton PeVatron, with the possible exception of the Galactic Centre \citep{2016Natur.531..476H,2018A&A...612A...9H,2020A&A...642A.190M}.
The association of a $\gamma$-ray source with a PeVatron can be indicated by the  absence of an exponential cut-off in the $\gamma$-ray spectrum below 100~TeV, as expected when CRs are accelerated up to PeV energies and produce $\gamma$-rays interacting with the ambient material. 

The source MGRO~J1908+06 is one of the best Galactic PeVatron candidates, thanks to its hard spectrum reaching energies above 100~TeV as observed by HAWC \citep{2020PhRvL.124b1102A}, and with no evidence of a cut-off. It was discovered by the MILAGRO collaboration \citep{2007ApJ...664L..91A} and later confirmed with the  HESS atmospheric Cherenkov telescope \citep{2009A&A...499..723A}. These authors reported the source detection  above 300~GeV, with a large angular size ($\sigma=0.34$\textdegree) and a hard spectrum with a photon index of $2.1$. MGRO~J1908+06 was observed at very high energy also with VERITAS \citep{2014ApJ...787..166A}, which revealed an extended morphology  ($\sigma=0.44$\textdegree) with  three  peaks of emission  
 and a spectrum with  photon index of $2.2$.
Observations of this sky region with HAWC  revealed emission  up to 100~TeV  spatially consistent with the VERITAS error box of MGRO~J1908+06, indicating the presence of very energetic particles \citep{2020PhRvL.124b1102A}.
Due to its complex spatial  structure,  difficult to study with the limited angular resolution of current instruments, the origin of the $\gamma$-ray emission from MGRO~J1908+06 is still uncertain. As discussed below, a few counterparts (the supernova remnant (SNR) G40.5-0.5, dense molecular clouds illuminated by cosmic rays from the nearby SNR,  and the pulsar PSR~J1907+0602) are compatible with the $\gamma$-ray source error box ($\sim 0.5$\textdegree), preventing a secure identification of this extreme accelerator and making it difficult to distinguish between  hadronic or leptonic interpretations of its emission.
This source has been studied also at radio wavelength by \cite{2020MNRAS.491.5732D}. These authors reported the presence of molecular material toward the SNR.\\
In this work we investigate the molecular cloud environment of MGRO~J1908+06 using data from molecular line radio surveys, modelling   its multiwavelength spectral energy distribution using published TeV spectra and our re-analysis of  \textit{Fermi-LAT} and \textit{XMM-Newton} data.

\begin{figure*}
    \centering
    \includegraphics[scale=0.6]{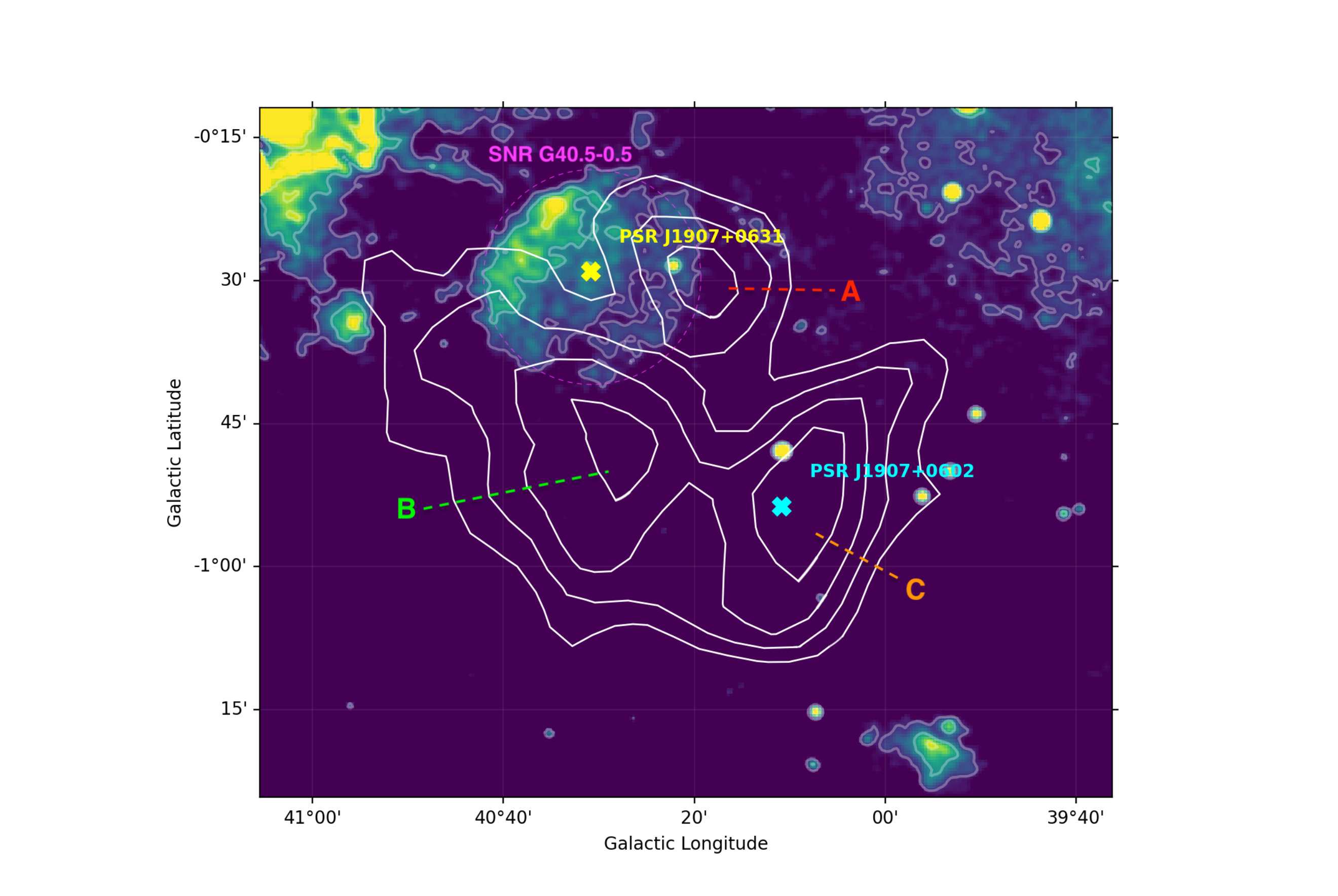}
    \caption{VLA Galactic Plane Survey at 1.4~GHz  \protect\citep{Stil_2006} map covering the entire source MGRO~J1908+06. 
    The white solid lines are the VHE contours taken from VERITAS \protect\citep{2014ApJ...787..166A}, where the significance level ranges from 3 to 5.2. The yellow and cyan crosses indicate the positions of PSR~J1907+0631 and PSR~J1907+0602, respectively. The thin dashed magenta line indicates the SNR~G40.5-0.5 shell, while the red, green and orange dashed lines mark the VERITAS emission lobes to which we refer in the following.}
    \label{map}
\end{figure*}

\section{Possible counterparts of MGRO~J1908+06 }
Several possible counterparts are present in the region of  MGRO~J1908+06: in Fig.~\ref{map} we show the continuum emission at 1.4~GHz obtained in the VLA Galactic Plane Survey (VGPS - \citep{Stil_2006}) with  superimposed the TeV contours obtained by VERITAS \citep{2014ApJ...787..166A}.  The radio data show the shell-like  SNR~G40.05-0.5, which is much brighter in the northern region. The positions of  the two pulsars PSR~J1907+0602 and PSR~J1907+0631 are also  indicated.

G40.05-0.5 is a middle aged supernova remnant, with estimated age between 20 and 40~kyr \citep{1980A&A....92...47D} and  uncertain distance. The $\Sigma - D$ relation gives a distance between 5.5 and 8.5~kpc \citep{1980A&A....92...47D}. A more accurate estimate can be obtained from the HI absorption spectrum (see \citealt{2017ApJ...843..119R} for details) but, as reported by \citet{2020MNRAS.491.5732D}, for   SNR~G40.5-0.5 this method resulted in a very noisy spectrum, probably due to the fact that the neutral gas is patchy.  
The relatively young and energetic pulsar PSR~J1907+0631 (characteristic age $\tau$=11~kyr, spin-down luminosity $\sim$5$\times10^{35}$~erg~s$^{-1}$) lies close to the centre of the SNR \citep{2017ApJ...834..137L}. 
Its dispersion measure (DM) implies a distance of 7.9~kpc \citep{2002astro.ph..7156C}, compatible with the range estimated for G40.05-0.5 and suggesting an association between these two objects.  
\citet{2020MNRAS.491.5732D} showed that, in principle, PSR~J1907+0631 could power the whole TeV source,  since this would require a conversion efficiency from rotational energy  to $\gamma$-rays of about 3\%,  in line with the efficiency   $\le 10\%$ of other known TeV sources associated to pulsar wind nebulae \citep{Gallant_2007}. However, this hypothesis is disfavoured by the pulsar position slightly outside the TeV contours and significantly offset from the centroid of the  $\gamma$-ray emission.

The $\gamma$-ray loud pulsar PSR~J1907+0602, discovered with \textit{Fermi-LAT} \citep{2010ApJ...711...64A}, is located in the southern
part of MGRO~J1908+06, slightly offset from the peak of the $\gamma$-ray excess counts.
This pulsar has a characteristic age of 19.5~kyr and a spin-down luminosity of $\sim$3$\times10^{36}$~erg~s$^{-1}$. The source distance was estimated to be 3.2~kpc \citep{2010ApJ...711...64A}, as derived from the DM of $\sim 82$~pc~cm$^{-3}$ with the electron distribution model of \cite{2002astro.ph..7156C}.

Another pulsar (PSR~J1905+0600, not marked in Fig.~\ref{map}) lies in this region,  but its  distance of $\sim$18 kpc and large characteristic age  of  6 Mys \citep{2004MNRAS.352.1439H} exclude it  as a possible counterpart.

In conclusion,  the two main candidates for the $\gamma$-ray emission from MGRO~J1908+06 are the SNR~G40.5-0.5, near the northern border of the TeV source,  and   PSR~J1907+0602, lying in the southern region.

\section{Data analysis and results}

\subsection{CO Analysis}
\label{sec:CO}

We have investigated the distribution of the CO gas in the environment of MGRO~J1908+06 in order to identify molecular material in spatial correlation with the SNR and the $\gamma$-ray emission. 
Any such association would be relevant for  hadronic models to explain the TeV emission and it would also provide some information on the source distance. 
We used the molecular line emission extracted from the FOREST Unbiased Galactic Plane Imaging (FUGIN) survey\footnote{Available   at     http://jvo.nao.ac.jp/portal/}. This project aims at  investigating the distribution, kinematics, and physical properties of both diffuse gas and dense molecular clouds in the Galaxy by observing simultaneously the $^{12}$CO, $^{13}$CO, and $^{18}$CO J=1-0 lines.  
This survey achieves the highest angular resolution to date ($\sim$20\arcsec)  for the Galactic plane, making it possible to find dense clumps located at farther distances than those seen in previous surveys. 

\begin{figure}
    \centering
    \includegraphics[width=.9\columnwidth]{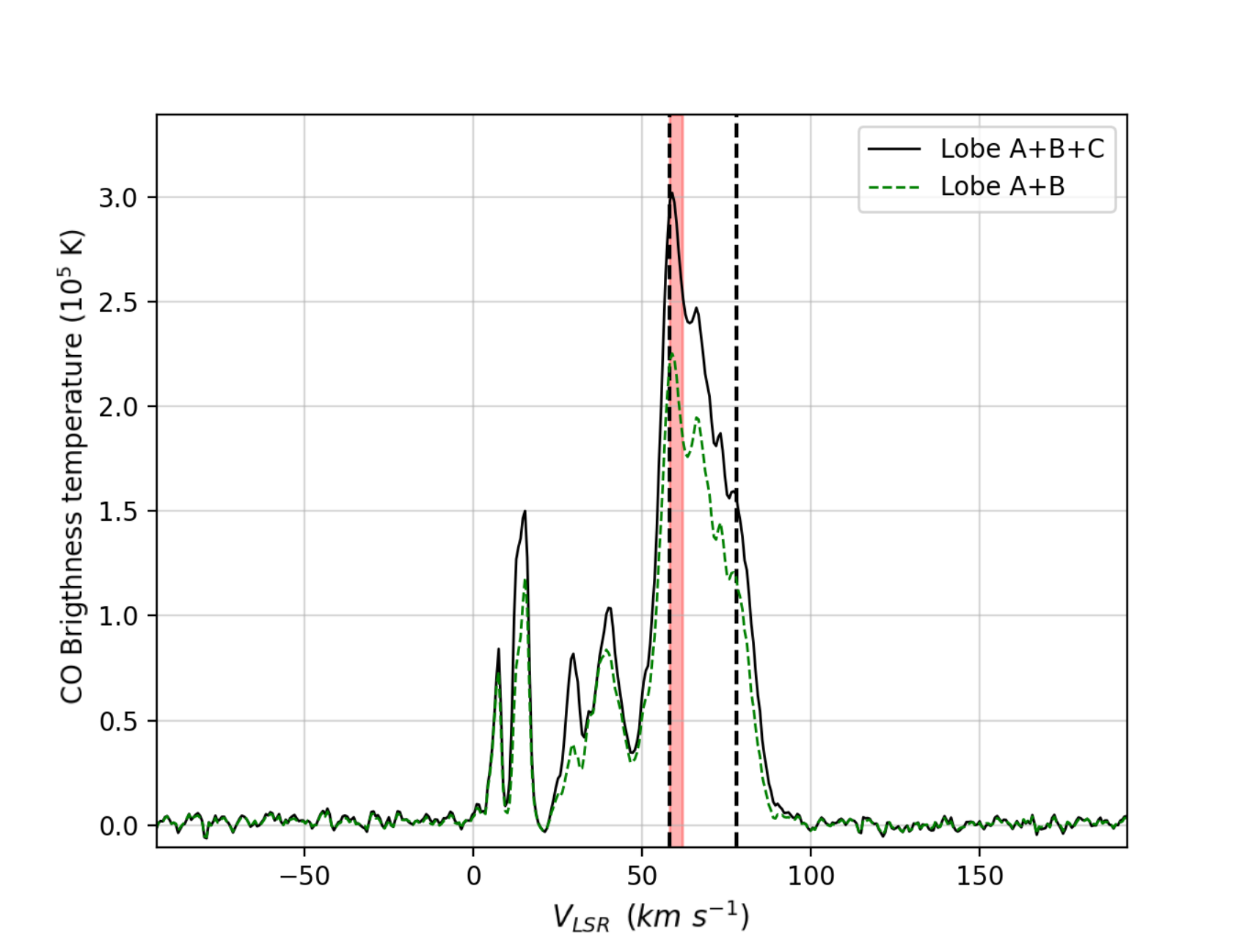}
    \includegraphics[width=.9\columnwidth]{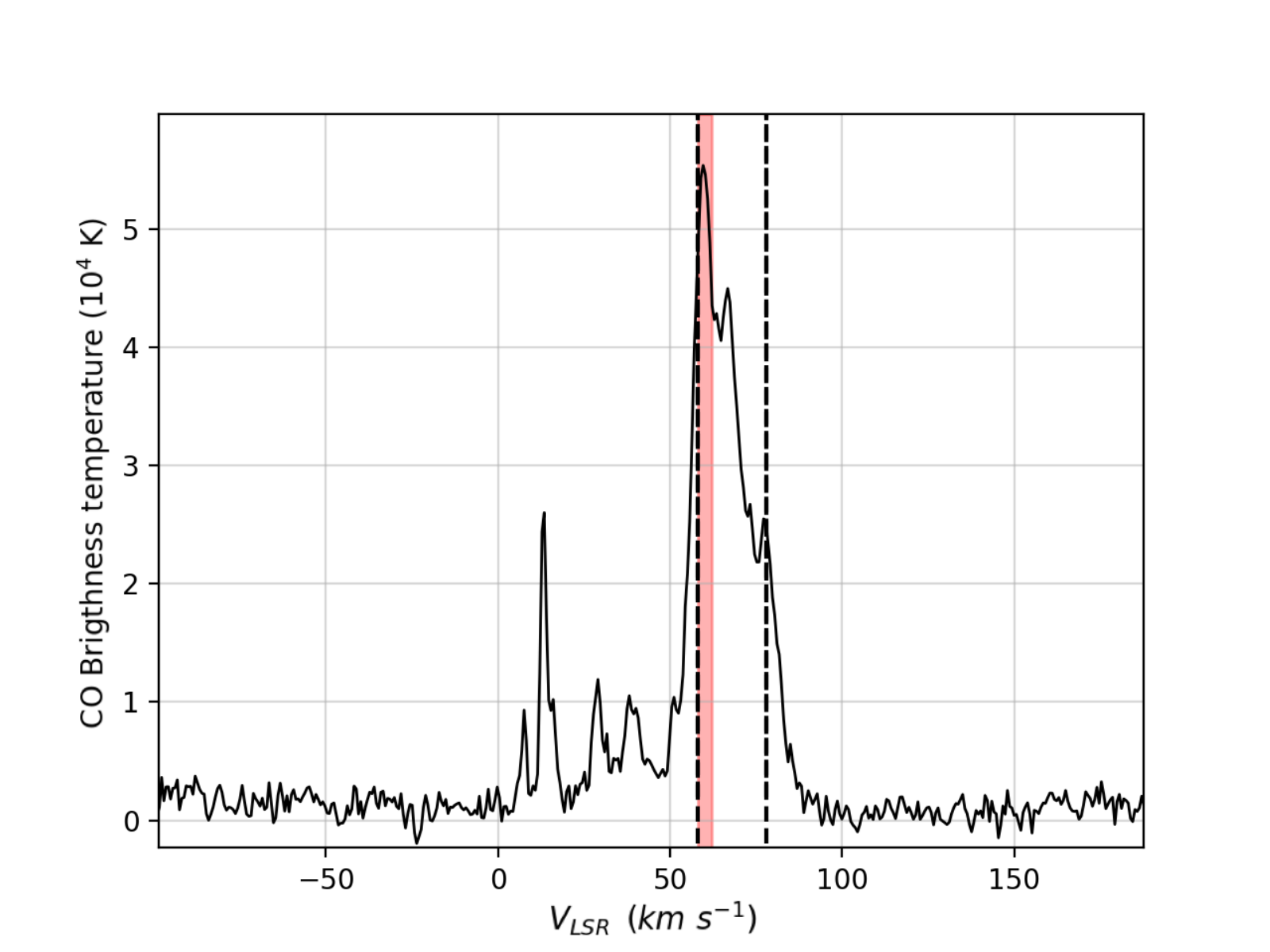}
    \caption{The  $^{12}$CO (top) and $^{13}$CO (bottom) summed spectra in the region of MGRO~J1908+06. The velocity interval between the two dashed lines (58--78~km~s$^{-1}$) represents the bulk of the emission, while the red zone marks the velocity range between 58 and 62~km~s$^{-1}$ (shown in Fig.~\ref{COmap}) that is the velocity range considered for the molecular cloud analysis (section~\ref{sec:CO}).}
    \label{fig:tb}
\end{figure}

We recovered the spectra in brightness temperature $T_{B}$ as a function of the local standard of rest velocity ($V_{LSR}$) for the whole region corresponding to the $3\sigma$ contours of the TeV emission, both in $^{12}$CO and $^{13}$CO. As shown in Fig.~\ref{fig:tb}, the bulk of the emission is concentrated between 50 and 80~km~s$^{-1}$.

\begin{figure*}
    \centering
    \includegraphics[scale=.6]{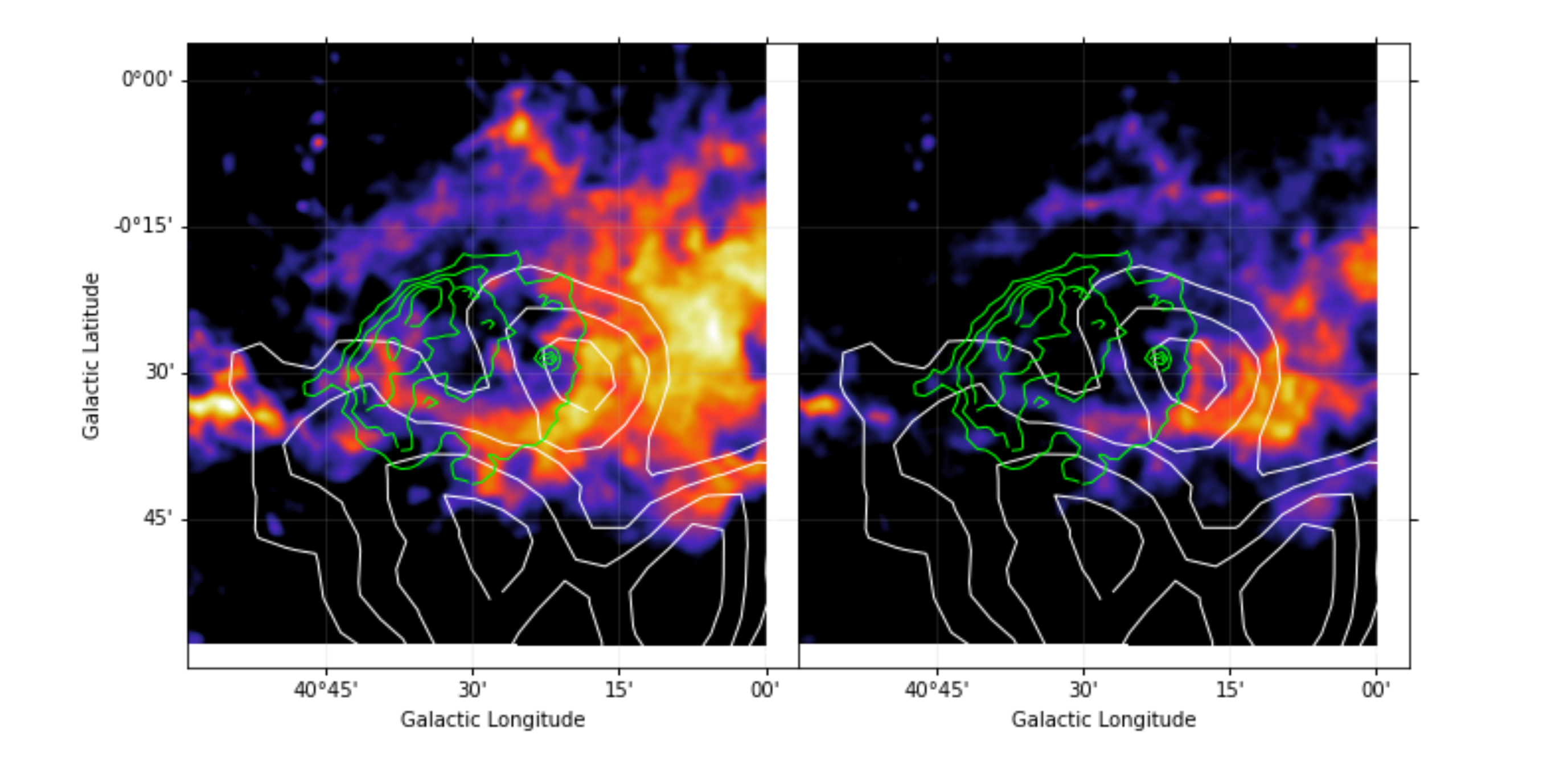}
    \caption{Maps of $^{12}$CO (left) and  $^{13}$CO  (right) emission in the MGRO~J1908+06 region  integrated between 58--62~km~s$^{-1}$.
    The white solid lines are the same as in Fig.~\ref{map}, while the green contours are the continuum emission from SNR~G40.5-0.5 at 1.4 GHz.}
    \label{COmap}
\end{figure*}

We plot in Fig.~\ref{COmap} the $^{12}$CO and $^{13}$CO molecular line emission integrated from 58 to 62~km~s$^{-1}$. The  contours presented in the figure are those of the VERITAS TeV emission \citep{2014ApJ...787..166A} and of the SNR~G40.5-0.5 at 1.4~GHz from the VGPS.
We denote the three maxima of $\gamma$-ray emission as lobes A, B, and C  (see Fig.~\ref{map}). 
The maps of Fig.~\ref{COmap} show  that   lobe A   overlaps with CO emission, lobe B   partially overlaps with CO emissions, while no obvious molecular clouds  association is seen for lobe C.

We concentrate on the molecular cloud in the 58--62~km~s$^{-1}$ velocity interval, as it overlaps both the A-B lobes and the  southern border of the SNR.
We obtain the distance of the   cloud     using the Galaxy rotation curve from \citet{1985ApJ...295..422C}, with $R_{\sun}$= 8.5~kpc and $v_{\sun}=220$~km~s$^{-1}$. The first Galactic quadrant presents distance ambiguity for positive radial velocities, so adopting 60~km~s$^{-1}$, we obtain near and far distances of 3.0 and 9.4~kpc, respectively.

To study the properties of the molecular gas, and in particular to estimate their density, we use the dendrogram technique \citep{2008ApJ...679.1338R}. A dendrogram is a topological representation of the significant local maxima in N-dimensional intensity data and the way these local maxima are connected along contours (or isosurfaces) of constant intensity. A local maximum, by definition, has a small region around it containing no data   greater than its value and, hence, a distinct isosurface containing only that local maximum can be drawn. The local maxima determines the top level of the dendrogram, which we refer to as the ``leaves'' , defined as the set of isosurfaces that contain a single local maximum.   
We identify and characterize molecular clouds in the CO data cube between 58--62~km~s$^{-1}$ using \textsc{astrodendro\footnote{Available at http://www.dendrograms.org/}}. This python algorithm efficiently constructs a dendrogram representation of all the emission in the selected region. 
The minimum value to consider  (any value lower than this will not be considered in the dendrogram) is set as the “detection level”, namely  5$~\sigma_{T}$, where $\sigma_{T}$ is the median RMS noise level in the dataset, so that only significant values are included in the dendrogram (T$_{\rm min}=3$~K).
Another consideration is about how significant a leaf has to be in order to be considered an independent entity. The significance of a leaf is measured from the difference between its peak flux and the value at which it is being merged into the tree. This parameter is set to  1 $\sigma_{T}$, which means that any leaf that is locally less than 1 $\sigma_{T}$ high is combined with its neighboring leaf (or branch) and is no longer considered as a separate entity.

Once an index of structures in the data has been produced by the algorithm,  it can be used to catalog the properties of each structure, such as integrated intensity, centroid position, spatial position angle, spatial extent, and spectral line-width.

We estimate the luminosity based on the zeroth moment, i.e., the sum of the intensity, and then translate the moments into estimates of physical quantities.
For these calculations, we consider the pixels in a cloud mask $\mathcal{M}$, i.e. only the pixels belonging to a single cloud identified by the segmentation algorithm. We measure the luminosity of each cloud as:
\begin{equation}
    L_{\rm CO}= A_{\rm pix} \Delta v  \sum_{i} T_{i}
\end{equation}
where $A_{\rm pix}$ is the projected physical area of a cube pixel in pc$^2$,
$\Delta v$ = 4~km~s$^{-1}$ is the channel width, and $T_{i}$ is the brightness of the cube pixels measured in K in the cloud mask $\mathcal{M}$.
We convert from luminosity to  mass,  scaling the extrapolated luminosity through the CO-to-H2 conversion factor, $\alpha_{\rm CO}$.
\begin{equation}
    M_{\rm CO}= L_{\rm CO}\alpha_{\rm CO}
\end{equation}
where we take $\alpha_{\rm CO}$= 4.35~$M_{\sun}$~pc$^{-2}$~(km~s$^{-1}$~K)$^{-1}$ at solar metallicity 
\citep{2013ARA&A..51..207B}. 
To measure cloud radii we convert from the deconvolved major and minor sizes, $\sigma_{\rm maj}$ and $\sigma_{\rm min}$, to a cloud radius measurement using:
\begin{equation}
    R= \eta \sqrt{\sigma_{\rm maj}\sigma_{\rm min}}  
\end{equation}
\noindent
The factor $\eta$  depends on the light or mass distribution within the cloud.  We adopt $\eta$=1.91 following \citet{2006PASP..118..590R}.
Our model approximates the cloud as a spherically symmetric object so that R also characterizes the object in three dimensions. Therefore, we do not apply any inclination corrections to R. The resulting mean cloud density is $\sim180$~particles~cm$^{-3}$ assuming a distance of 3~kpc, while it is  $\sim 60$~particles~cm$^{-3}$ assuming 9~kpc.
 
\subsection{\textit{Fermi-LAT} data analysis}\label{sec:G-ray}

We analyzed 12 years of \textit{Fermi-LAT} data, obtained from 2008-09-01 to 2020-12-16, exploiting the  Pass 8 data processing (P8R3) with the public \textsc{fermitools} (v2.0.0) and \textsc{fermipy} packages (v1.0.0). We selected the Pass 8 ‘source’ class and ‘front+back’ type events coming from zenith angles smaller than 90\textdegree \, and from a circular region of interest (ROI) with radius of 10\textdegree \, centered at   R.A. = 286.97\textdegree \, and Dec. = 6.03\textdegree \, (J2000). The instrument response function version P8R3-SOURCE-V3 was used. We selected only the events in the 10~GeV--1~TeV energy range, to avoid the contribution from PSR~J1907+0602 (see fig.~4 of \citealt{2010ApJ...711...64A}). We included in  the background model all the sources from the 4FGL catalog within the ROI, as well as the Galactic (gll-iem-v07.fits) and the isotropic (P8R3-SOURCE-V3-v1) diffuse components. 

We performed a binned analysis with five bins per energy decade and  spatial pixel size of 0.05\textdegree \ . In the maximum likelihood fitting, the normalization parameter of all the sources within 3\textdegree \, of the ROI centre, as well as the diffuse emission components, were left free to vary. Instead the parameters of all the other sources at more than 3\textdegree \, were fixed to the values given in the 4FGL catalog \citep{2020ApJS..247...33A}. To describe the  spatial morphology of MGRO~J1908+06, we used the VERITAS emission region at 3$\sigma$ level (i.e. the outermost contour in Fig.~\ref{map} ), while for the spectral model we assumed a power law with photon index $\Gamma = 1.6$. This leads to a detection significance $\sqrt{TS}\sim 6$ in the energy band considered. 
The $\gamma$-ray flux  was obtained by binning the $\gamma$-ray data in the range from 10 to 1000~GeV into four energy intervals, and performing a binned likelihood analysis in each energy bin. The resulting \textit{Fermi-LAT} spectral energy distribution is plotted in Fig.~\ref{fig:IC_Sed}.

\subsection{X-ray Analysis} \label{sec:X-ray}

To study the X-ray emission in the vicinity of PSR~J1907+0602 we used a 52~ks long observation carried out on 2010 April 26 with the \textit{XMM-Newton} satellite. We analyzed the data of the EPIC-MOS instrument that was operated in full frame imaging mode and with the medium thickness optical filter. We excluded time intervals with high background, resulting in net exposure times of 36 and 38~ks for the two MOS cameras. 

Using the Extended Source Analysis Software  (ESAS\footnote{https://heasarc.gsfc.nasa.gov/docs/xmm/esas/cookbook/xmm-esas.html}), we extracted the spectra from a circular region of 5 arcmin radius centered at the position of PSR~J1907+0602 (excluding a circle of 30\arcsec radius around the source) and from a concentric annular region  with radii 5 and 12.5 arcmin. The latter was used to estimate the X-ray background (which in this sky region is dominated by the Galactic Ridge diffuse emission). Comparison of the two spectra showed no evidence for diffuse emission associated with PSR~J1907+0602,  with an upper limit (at 95\% c.l.) of 1.2$\times10^{-15}$~erg~cm$^{-2}$ s$^{-1}$~arcmin$^{-2}$ on the surface brightness in the 1--10~keV energy range.

\section{Origin of the $\gamma$-ray emission}

Emission at TeV energies indicates the presence of ultra-relativistic particles which, in principle,  can  produce it  through Inverse Compton (IC) scattering of the CMB, IR and/or star-light seed photons by electrons, or through the decay of neutral pions resulting from proton-proton (and/or other nuclei) interactions. 
In this Section, we first explore the possibility that a single mechanism is responsible for the  emission from the whole trilobed region in either the leptonic or hadronic scenario.

We then consider the possibility of a two-zone model, 
in which both components (hadronic and leptonic) are present.  
This scenario can be originated by a source able to efficiently accelerate both protons and electrons, or by 2 sources lying in the same sky region. 

In the following, we use the \textit{Fermi-LAT} and \textit{XMM-Newton} results derived as described above and the TeV spectra  obtained with VERITAS \citep{2014ApJ...787..166A} and MILAGRO \citep{2007ApJ...664L..91A}. 
We have not used the HAWC data reported in \citep{2020PhRvL.124b1102A} for our fit, 
because they are inconsistent with both the VERITAS and HESS data in an energy range (1-10~TeV) where these instruments proved to show reliable results.
Also HESS data were not considered in the fitting procedure as they are fully compatible with the VERITAS ones.
This discrepancy is explained in \citep{2020PhRvL.124b1102A} as consequence a the larger source extent observed by HAWC.
However, we note that our model reproduces quite well the slope of the HAWC spectrum at energies of around 100~TeV, hence in the range where there is consistency between the results reported in \citep{2020PhRvL.124b1102A} and \citep{Abeysekara_2017}.

The X-ray upper limit, derived for a region of about 75 arcmin$^2$, was rescaled for the larger area enclosed in the  3$\sigma$ contours of the  TeV emission ($\sim$ 1400 arcmin$^2$). This is a conservative assumption, because any diffuse X-ray emission produced by particles accelerated by the pulsar would likely decrease in intensity at larger distances.

\subsection{1-zone leptonic Hypothesis} \label{sec:1lep}

In the leptonic scenario, we  assume that the whole emission from MGRO J1908+06 is due to a population of relativistic electrons interacting, through IC,  with the ISRF photons.
These electrons can be supplied by the energetic pulsar PSR~J1907+0602 or by the SNR~G40.5-0.5.

In order to infer the  properties of the parent particle distribution, we fit the multiwavelength SED through a Markov Chain Monte Carlo (MCMC) procedure using the \textsc{naima} python package (see \citealt{2015ICRC...34..922Z} for a detailed description of the fitting procedure and of the IC radiative model). We modeled the  three dominant photon fields with energy densities fixed at  
$\epsilon_{\rm CMB}$ = 0.261~erg~cm$^{-3}$ for the Cosmic Microwave Background, 
$\epsilon_{\rm FIR}$ = 0.5~erg~cm$^{-3}$ for the far-infrared dust emission, and $\epsilon_{\rm NIR}$ = 1.0~erg~cm$^{-3}$ for the near-infrared stellar emission.
The electron distribution was modeled as a broken power law with an exponential cut-off with all the parameters free to vary during the fitting procedure. 

\begin{figure}
    \centering
    \includegraphics[width=\columnwidth]{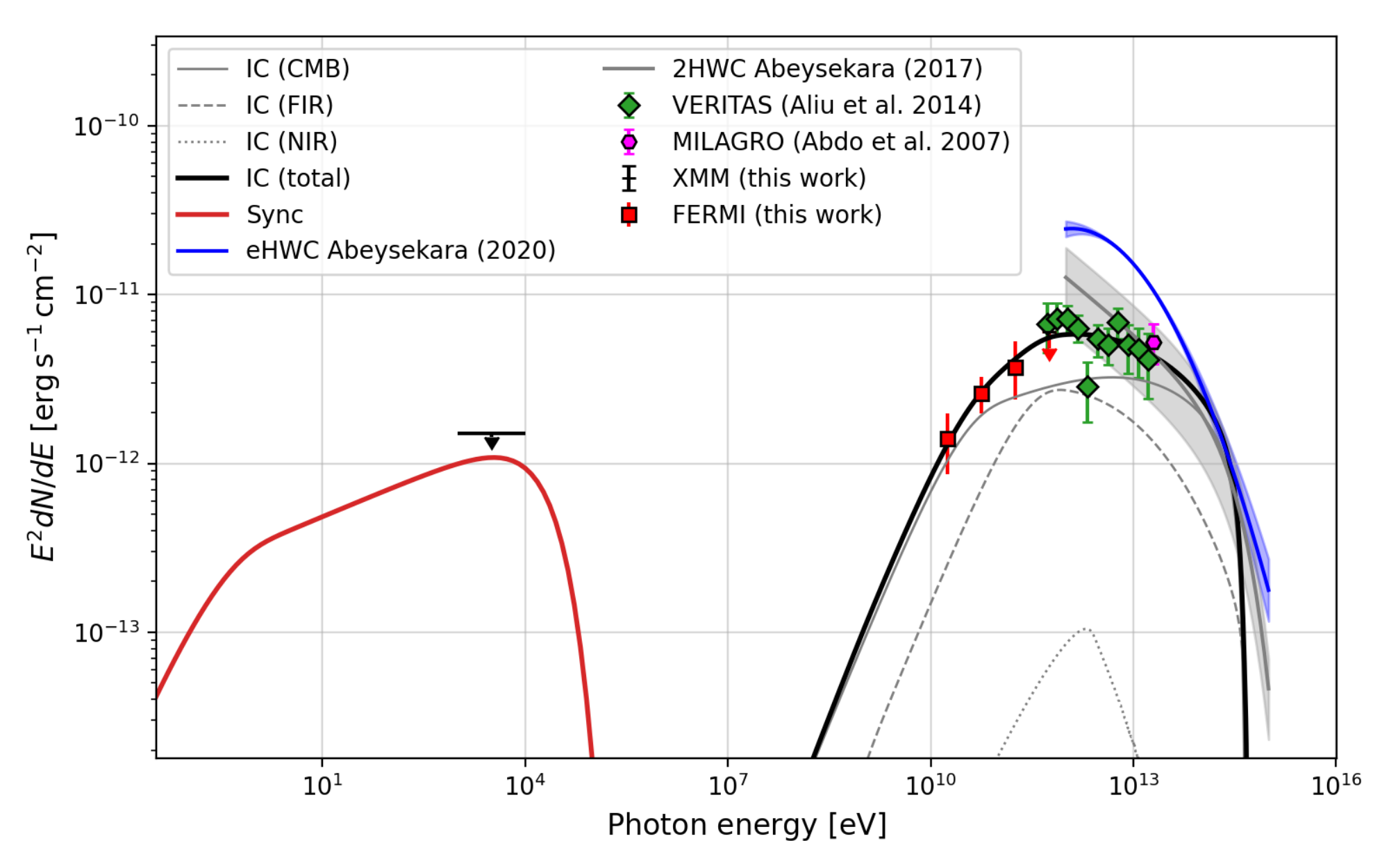}
    \caption{Multiwavelength SED of MGRO~J1908+06 fitted with  the leptonic scenario (Inverse Compton + synchrotron) with B=1.2~$\mu$G. The black line shows the   total IC emission,  whose components are indicated by   the thin solid, dashed and dotted black line for the CMB, FIR and NIR, respectively. The red solid line is for the synchrotron emission. The \textit{XMM-Newton} upper limit (black arrow) as well as the \textit{Fermi-LAT} data (red squares) are from this work. The other data are taken from \protect\cite{2014ApJ...787..166A} (VERITAS - green diamonds) and \protect\cite{2007ApJ...664L..91A} (MILAGRO - pink hexagons). The blue and gray butterflies are the HAWC models  from \protect\cite{2020PhRvL.124b1102A} and from \protect\cite{Abeysekara_2017}, plotted for comparison, but not used in the fit. }
    \label{fig:IC_Sed}
\end{figure}

We obtained a best fit (see Fig.~\ref{fig:IC_Sed}) with the parameters for the electron distribution shown in Table~\ref{tab:par}.  

The high-energy electrons responsible for the TeV emission interact also with the ambient magnetic field producing synchrotron radiation. We found that the electron population obtained in our best fit can be reconciled with the \textit{XMM-Newton} upper limit in the X-ray band only for ambient magnetic fields (B) smaller than 1.2~$\mu$G.  

\begin{figure}
    \centering
    \includegraphics[width=\columnwidth]{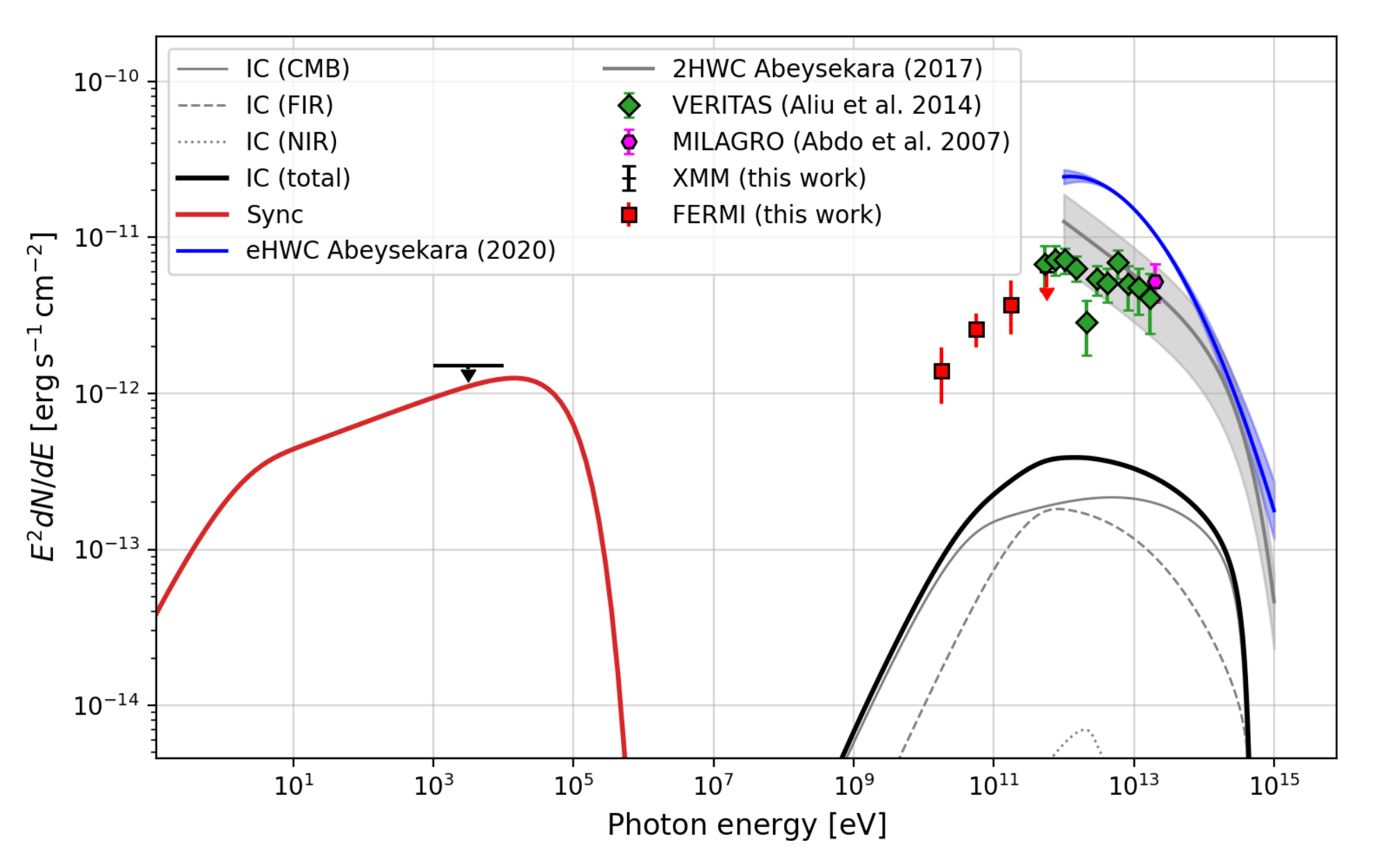}
    \caption{
    Same as Fig.~\ref{fig:IC_Sed}, but with B=5~$\mu$G and the electron population normalization reduced by a factor  15.
    }
    \label{fig:IC_B}.
\end{figure}

This limit is rather  small compared to the typical value of the Galactic magnetic field of 5~$\mu$G \citep{2015ASSL..407..483H}. 
If we assume a magnetic field of 5~$\mu$G and the same spectral slope of the best fit, the normalization of the electrons population must be reduced by a factor $\sim$15 to be consistent with the X-ray upper limit (see Fig.~\ref{fig:IC_B}).
This results shows that a 1-zone leptonic model alone cannot explain the whole multiwavelength set of data.


\subsection{1-zone hadronic hypothesis} \label{sec:HH}
 
\begin{figure}
    \centering
    \includegraphics[width=\columnwidth]{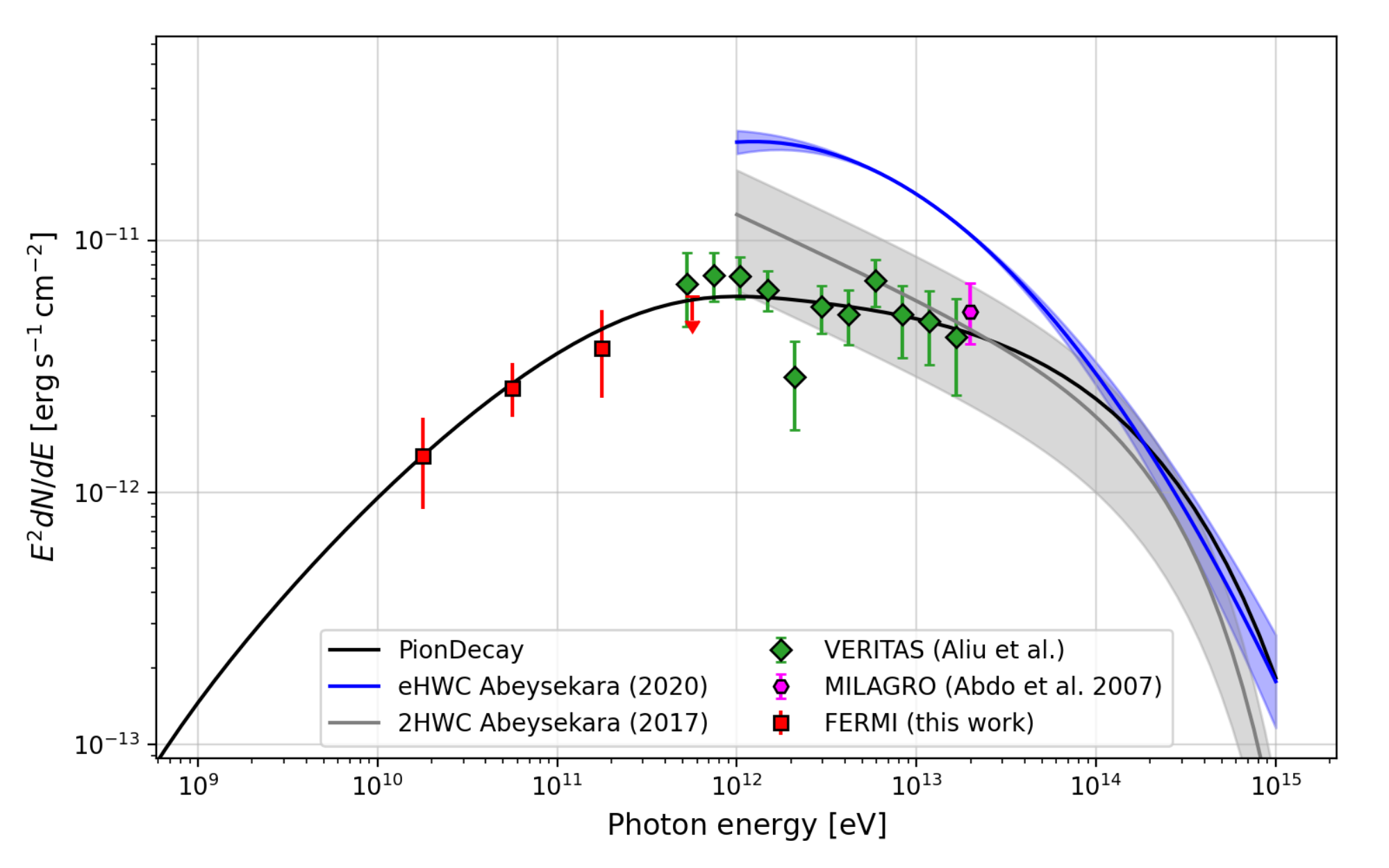}
    \caption{MGRO~J1908+06 hadronic emission model. The black line shows the Pion Decay best fit model. Data are the same as in Fig.~\ref{fig:IC_Sed}   }
    \label{fig:pion}
\end{figure}

In the hadronic scenario we assume that a population of relativistic protons interacts with dense interstellar material and produces TeV photons via pion decay.
A good candidate for the acceleration of these protons is the SNR~G40.5-0.5. In fact, our analysis of the molecular gas around the SNR demonstrated the presence of molecular clouds in good spatial correlation with the SNR shell. We used \textsc{naima} to fit the $\gamma$-ray SED, assuming a proton distribution described by a broken power law with exponential cut-off and 
pion decay as radiative model. The parameters of the best fit, shown in Fig.~\ref{fig:pion}, are given in Table~\ref{tab:par}.

With the average densities of the clouds derived in section~\ref{sec:CO}, $180$ or $60$ particles~cm$^{-3}$ depending on the considered distance, the total proton energy required by the fit is $7\times10^{47}$~erg or $2\times10^{49}$~erg, respectively. 

\subsection{2-zone model}\label{2z}

Considering the complex spatial distribution of the VHE emission, it can not be excluded that both a TeV PWN powered by PSR~J1907+0602 and hadronic processes associated to the SNR~G40.5-0.5 contribute to the $\gamma$-ray emission observed from this sky region. Therefore, we have also explored a hybrid emission model in which the TeV emission is due to the superposition of leptonic and hadronic components from these two sources.
Of course, in this scenario, MGRO~J1908+06 might consist of two physically separated sources, not necessarily at the same distance. 
\begin{figure}
    \centering
    \includegraphics[width=\columnwidth]{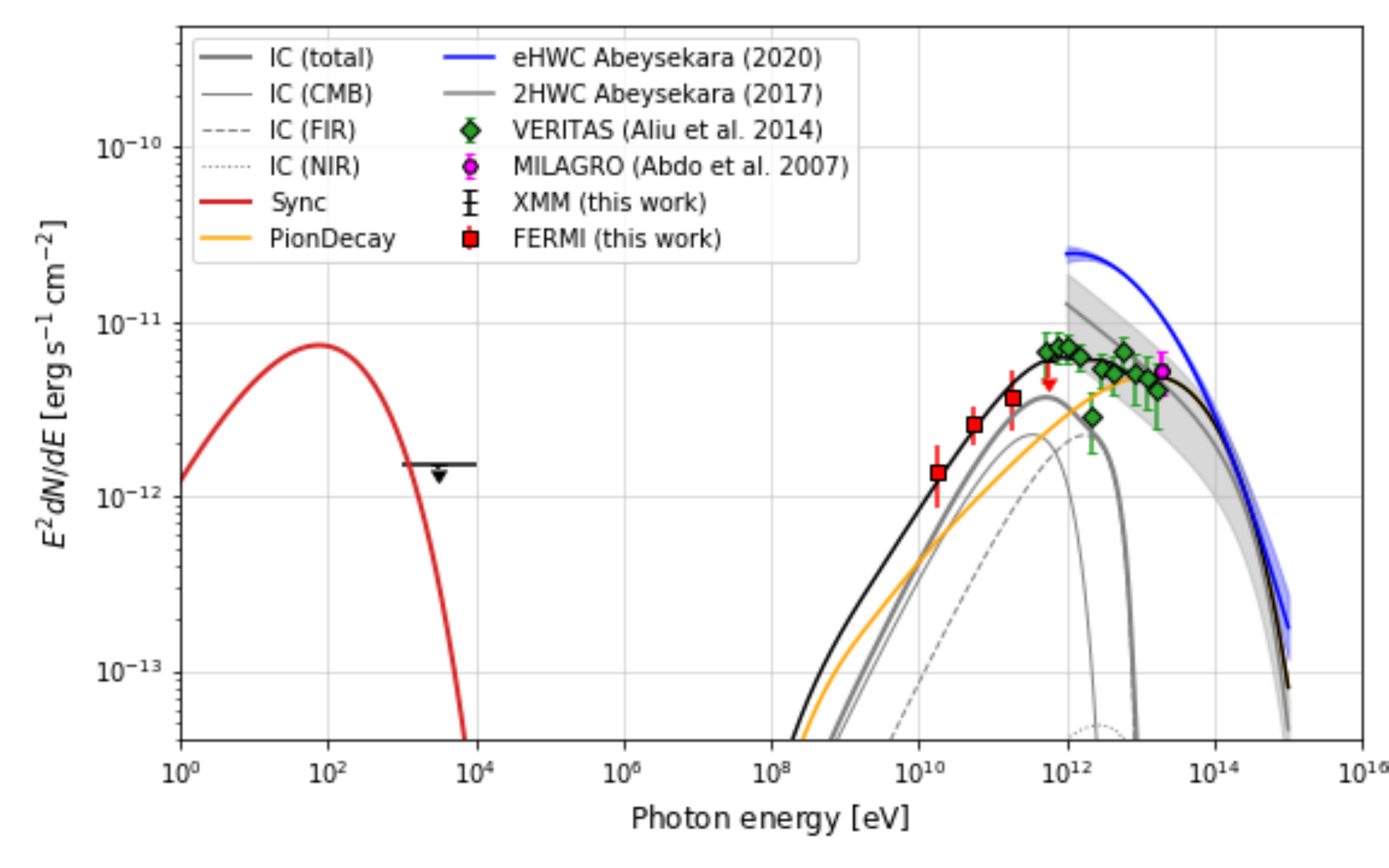}
    \caption{MGRO~J1908+06 2-zone emission model. The black line shows the total emission while the gray lines and the orange lines are for the IC and pion decay emission respectively. Other colors are the same as in Fig.~\ref{fig:IC_Sed}. The magnetic field is B=4~$\mu$G. }
    \label{fig:2zone}
\end{figure}

To fit the spectral energy distribution we assume that the steep spectrum at GeV energies has a leptonic origin, while the hadronic emission is responsible for the softer part at TeV energies. We used \textsc{naima} to recover the radiative models from a particle distribution. We used an exponential cut-off broken power law both for electrons and protons.
The resulting emission model is plotted in Fig.~\ref{fig:2zone}, while the assumed  parameters are reported in Table~\ref{tab:par}.
The recovered total protons energy for distances of 3 and 9~kpc is $4\times10^{47}$~erg and $10^{49}$~erg,  respectively, while the electrons energy is $9\times10^{46}$~erg.

\begin{table*}
    \centering
    \begin{tabular}{c|c|c|c|c|c|c|c|c}
    \hline
    \hline
         Model & Component & d &$ \Gamma_1$ & $\Gamma_2$& $W$ & $E_{0}$&$E_b$ & $E_{c}$ \\
                &            &   (kpc) &       &  &    (erg) & (TeV) & (TeV) & (PeV) \\
                \hline
        1-zone & Leptonic & 3 & 1.0 $\pm$ 0.4 & 2.6 $\pm$ 0.1& 2$\times10^{47}$  & 10& 2.7 $\pm$ 0.7 & 7.1 $\pm$ 6.0 \\ 
        \hline
        1-zone & Hadronic & 3 & 1.0 $\pm$ 0.1 &2.1 $\pm$ 0.1 & 7$\times10^{47}$  & 30 & 2.8 $\pm$ 0.8   & 3.0 $\pm$ 0.9 \\ 
        \hline
        1-zone & Hadronic & 9 & 1.1 $\pm$ 0.2 &2.1 $\pm$ 0.1 & 2$\times10^{49}$  & 30 & 3.4 $\pm$ 1.2   & 1.9 $\pm$ 0.5 \\ 
        \hline
        2-component & Leptonic & 3 &1.2 & 1.2 & 9$\times 10^{46}$ & 10 & 0.2 & 0.011 \\
        & Hadronic & 3 &1.6 & 2.0 & 4$\times 10^{47}$ & 30 & 200 &>1\\
        & Hadronic & 9 &1.6 & 2.0 & 1$\times 10^{49}$ & 30 & 200 & >1\\
         \hline
    \end{tabular}
    \caption{Parameters for all emission models considered.  $\Gamma_1$ and $\Gamma_2$ are the indices before and after the break ($E_b$), W is the particles total energy, $E_0$ is the reference energy while $E_c$ is the cut-off energy. }
    \label{tab:par}
\end{table*}

\section{Discussion}

The association of MGRO~J1908+06 with an offset relic PWN driven by PSR~J1907+0602 was initially considered the most likely origin of the VHE \citep{2010ApJ...711...64A}. Our results show that,  for reasonable values of the ambient magnetic field,  a leptonic emission model fitting the $\gamma$-ray spectrum from the whole source would produce a synchrotron  X-ray flux incompatible with the upper limit in the few keV region. 
The leptonic interpretation is also disfavored by the spatial shape of the TeV emission, extending far from the pulsar position and without evident sign of spectral softening with distance from the pulsar, as it would be expected from electron cooling \citep{2014ApJ...787..166A}. 
We further note that, due to the Klein-Nishina suppression of the IC cross section at high energies, a rather large value of the electron maximum energy is required to fit the $\gamma$-ray spectrum.
 
Our analysis of the molecular gas around SNR~G40.5-0.5 demonstrates the presence of molecular clouds in good spatial correlation with the SNR (see also \citealt{2020MNRAS.491.5732D}) and motivates the exploration of a hadronic scenario. 
The cloud densities required by the best fit proton distribution used to reproduce the observed $\gamma$-ray spectrum are consistent with the ones that we derived from an analysis of the CO data, independent of the source association with the near (3~kpc) or far  (9~kpc)  source distance. However, the VHE emission of MGRO~J1908+06 extends beyond the  spatial   distribution of the target material, with  no obvious molecular cloud counterparts in  the southern region. The very hard photon index $\Gamma_1$  required to fit the \textit{Fermi-LAT} data (not seen in other TeV sources associated with SNRs) also disfavors a fully hadronic model for MGRO~J1908+06.
 
The difficulties discussed above for single zone models are easily solved assuming that the MGRO~J1908+06 spectrum is a sum of two components (hadronic and leptonic). 
The available data do not allow us to perform a spatially-resolved spectral analysis. Thus we cannot  exclude that these two components arise from different zones of the source (for example, a TeV PWN powered by PSR~J1907+0602 could be responsible for the southern lobe, and the interaction between the SNR~G40.5-0.5 and the molecular clouds for the northern part).

An important result of this analysis is that for both the 1-zone hadronic model and the 2-component model the maximum energy of the emitting protons required to fit the gamma-ray spectrum is greater than 1~PeV.

\section{Conclusions}

Our multiwavelength modeling of MGRO~J1908+06 confirms that this source is one of the best Galactic PeVatron candidates. 
We found that single-zone models, although in principle justified by the presence of plausible counterparts in both the leptonic and hadronic scenarios (a pulsar wind nebula powered by PSR~J1907+0602 or SNR~G40.05-0.5 interacting with molecular clouds, respectively), run into problems to explain the multiwavelength and spatial morphology properties of MGRO~J1908+06.
Therefore a 2-zone model is preferred to describe the emission from this source.
Spatially resolved data, as those that  will be provided by the next generation of Cherenkov telescopes such as the upcoming ASTRI Mini-array and the Cherenkov Telescope Array (CTA), are needed to separate the emission components of this source.

\section*{Acknowledgements}
This work made use of data from FUGIN, FOREST Unbiased Galactic Plane Imaging survey with the Nobeyama  45-m telescope, a legacy project in the Nobeyama 45-m radio telescope. 

\section*{DATA Availability}
The data underlying this article will be shared on reasonable request to the corresponding author.
\

%%%%%%%%%%%%%%%%%%%% REFERENCES %%%%%%%%%%%%%%%%%%

\bibliographystyle{mnras}
\bibliography{example} 
\bsp
\label{lastpage}
\end{document}